\documentclass[10pt,conference,letterpaper]{IEEEtran}
\IEEEoverridecommandlockouts
\usepackage{amsmath}
\usepackage{svg}
\usepackage{mathtools}
\usepackage{subfigure}  
\usepackage{multirow}
\usepackage{arydshln}
\usepackage{graphicx}
\usepackage{pythonhighlight}
\usepackage{epstopdf}
\usepackage{color}
\usepackage{amssymb}
\usepackage{balance}
\usepackage{url}
\usepackage{amssymb}
\usepackage{multirow}
\usepackage{arydshln}
\usepackage{threeparttable}
\usepackage{color}
\usepackage{algorithm}
\usepackage{algorithmic}
\usepackage{bm}
\usepackage{mathrsfs}
\usepackage{caption}
\usepackage{mathtools}
\usepackage{textcomp}
\usepackage{xcolor}
\usepackage{fancyhdr}
\usepackage{booktabs}
\usepackage{graphicx}
\usepackage{authblk}
\usepackage{multirow}

\def\BibTeX{{\rm B\kern-.05em{\sc i\kern-.025em b}\kern-.08em
    T\kern-.1667em\lower.7ex\hbox{E}\kern-.125emX}}

\title{A Time- and Energy-Efficient CNN with Dense Connections on Memristor-Based Chips}
\author[1]{Wenyong Zhou*}
\author[1]{Yuan Ren*}
\author[1]{Jiajun Zhou}
\author[2]{Tianshu Hou}
\author[1]{Ngai Wong*}
\affil[1]{Department of Electrical and Electronic Engineering, The University of Hong Kong, Hong Kong}
\affil[2]{Department of Micro/Nano Electronics, Shanghai Jiao Tong University, Shanghai, China}

\begin{document}
\renewcommand\arraystretch{1.5}
\maketitle
\thispagestyle{empty}
\pagestyle{empty}

\begin{abstract}
Designing lightweight convolutional neural network (CNN) models is an active research area in edge AI. Compute-in-memory (CIM) provides a new computing paradigm to alleviate time and energy consumption caused by data transfer in von Neumann architecture. Among competing alternatives, resistive random-access memory (RRAM) is a promising CIM device owing to its reliability and multi-bit programmability. However, classical lightweight designs such as depthwise convolution incurs under-utilization of RRAM crossbars restricted by their inherently dense weight-to-RRAM cell mapping. To build an RRAM-friendly yet efficient CNN, we evaluate the hardware cost of DenseNet which maintains a high accuracy vs other CNNs at a small parameter count. Observing the linearly increasing channels in DenseNet leads to a low crossbar utilization and causes large latency and energy consumption, we propose a scheme that concatenates feature maps of front layers to form the input of the last layer in each stage. Experiments show that our proposed model consumes less time and energy than conventional ResNet and DenseNet, while producing competitive accuracy on CIFAR and ImageNet datasets.
\end{abstract}

\section{Introduction}
\label{sec:introduction}
Deep neural networks (DNNs)~\cite{ResNet} have shown impressive capabilities in numerous tasks. In the area of computer vision, convolutional neural networks (CNNs)~\cite{ResNet,DenseNet, mobilenet} have gained popularity since more than a decade ago. In addition to expanding their representation power, the design of lightweight CNNs~\cite{mobilenet,shufflenet} has attracted increased attention due to the proliferation of edge AI. Some works attempt to compress a CNN by pruning~\cite{deepcompression}, quantization~\cite{quant} or knowledge distillation~\cite{distilling}. Other works try to re-architect for efficient networks~\cite{mobilenet,shufflenet}.

In addition to reducing the number of parameters (i.e., weights), hardware optimization is critical in reducing latency and energy consumption when deploying modern CNNs on edge devices. Most compute platforms adopt a von Neumann architecture which separates the compute and memory units. However, the frequent data transfer between the two units leads to a wastage of energy up to 60\%+~\cite{benchmark}. Compute-in-memory (CIM)~\cite{rram}, as its name suggests, conducts computation right at the storage, thereby alleviating/bypassing this memory wall hurdle. CIM chips can be realized via the emerging non-volatile memories (NVMs), in which the resistive random-access memory (RRAM)~\cite{rram} demonstrates significant potential due to its reliability and multi-bit programmability. In particular, the crossbar array is widely adopted in RRAM chips to accelerate the inference of neural networks. However, such crossbar structure hinders the direct porting of CNN design optimized for graphical processing units (GPUs), such as MobileNet~\cite{mobilenet}. Moreover, efficient network designs targeting smaller model sizes and fewer floating point operations (FLOPs), such as ShuffleNet~\cite{shufflenet}, do not readily warrant low hardware consumption on RRAM crossbars. Besides chip area, latency and energy consumption are important hardware measures when deploying models onto RRAM. Subsequently, RRAM-friendly CNNs that are efficient in time and energy consumption constitute an important research topic. To this end, this paper proposes a time- and energy-efficient RRAM-friendly CNN by preserving the merit of dense connections while novelly balancing time and efficiency.

Henceforth, we first discuss the inefficiency of depthwise convolution on crossbars, restricted mainly by the weight-to-RRAM mapping. Next we simulate the hardware cost of DenseNet~\cite{DenseNet} on RRAM using NeuroSim~\cite{neurosim}. It is found that the linearly increasing inputs and transition modules lead to low crossbar utilization, causing a major waste of time and energy. Based on these analyses, we then propose a new RRAM-friendly CNN that concatenates feature maps of initial layers to construct the input for the last layer in each stage. Experiments show that our modified architecture consumes lower time and energy than standard ResNet~\cite{ResNet} and DenseNet, while achieving a highly competitive performance on CIFAR and ImageNet datasets.

\section{Crossbar-Oriented Hardware Cost Analysis}
\label{sec:rethink}
\begin{figure}[!t]
\centering
\includegraphics[scale=0.26]{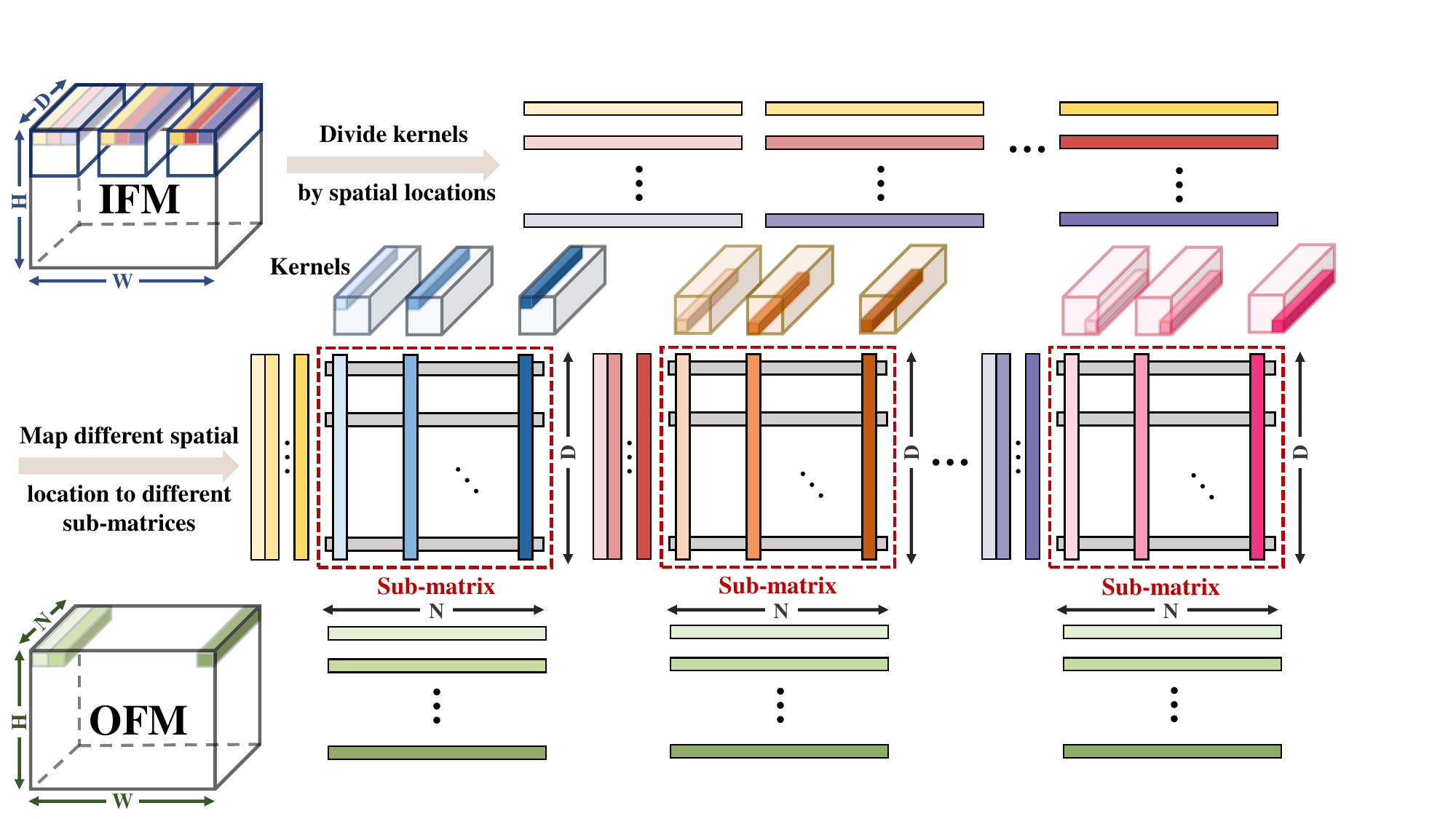}
\caption{The mapping in NeuroSim~\cite{neurosim} maps the weights across the spatial domain to a group of sub-matrices (IFM: input feature map, OFM: output feature map).}
\label{fig:mapping}
\vspace{-0.3cm}
\end{figure}

An RRAM-based crossbar array implements dot-product computation in a parallel fashion by summing currents along each column concurrently. Traditional CIM designs unroll each 3D kernel of the convolutional layers into a vertical column of a large weight matrix, where the input data will be accessed multiple times, whereby much energy is consumed in interconnects and buffers. NeuroSim~\cite{neurosim}, depicted in Fig.~\ref{fig:mapping}, divides convolution kernels and assigns the input data into different processing elements (PEs) according to their spatial locations to maximize both weight and input data reuse. Therefore, $k\times k$ sub-matrices are needed for $k\times k$ kernels and the input data at each kernel spatial location are sent to the corresponding sub-matrix.

\paragraph{Inefficiency of depthwise convolution} 
Popular efficient networks, such as MobileNet and ShuffleNet, reduce the model sizes and FLOPs using depthwise convolution and $1\times 1$ (viz. pointwise) convolution bottleneck architecture. Although $1\times 1$ convolution suits the mapping method, depthwise convolution is inefficient when deployed on crossbar arrays. Referring to Fig.~\ref{fig:mapping}, the input of each crossbar sub-matrix consists of flattened vectors whose entries are located in the same spatial position but across different channels in the input feature maps (IFMs). Since depthwise convolutions take place in each channel separately, the input vectors have to be padded with zeros in other channels to meet the mapping requirement. This induces not only redundant chip areas but also substantial time and energy consumption. Therefore, splitting standard $3\times 3$ convolution into depthwise and pointwise convolutions (aka separable convolution) leads to additional hardware resources instead of conservation.

\begin{figure}[!t]
\centering
\includegraphics[scale=0.38]{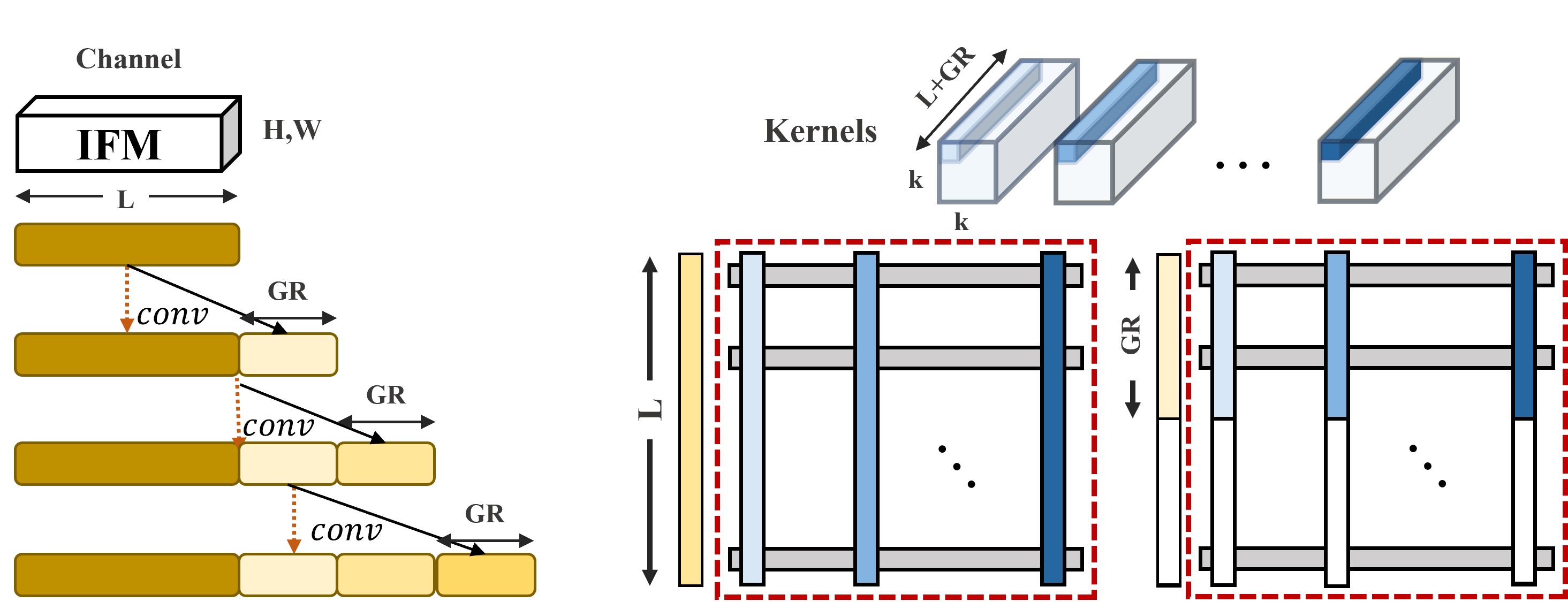}
\caption{(Left) Linearly increasing input channels in DenseNet. (Right) Mapping the 2nd layer on the left to RRAM crossbars. Blanked tails denote unused RRAM cells.}
\label{fig:densenet}
\end{figure}

\paragraph{Inefficiency of DenseNet} In view of depthwise convolution being RRAM-incompatible, DenseNet appears to be a potential candidate for RRAM-friendly architecture since it adopts $3\times 3$ and $1\times 1$ convolution kernels, and performs well with a relatively small number of weight parameters. As a morph of ResNet, DenseNet connects every two layers in each stage. Unlike ResNet aggregating features from shallower layers by summation, DenseNet concatenates features from all previous layers along the channel dimension (cf. Fig.~\ref{fig:densenet}), thus preserving all information in its original form. The growth rate (GR) controls the number of feature maps being passed from one layer to all subsequent layers. Nonetheless, concatenating features of all previous layers in the channel dimension leads to linearly increasing channels. As seen in Fig.~\ref{fig:mainResult} of Section~\ref{sec:experiments}, DenseNet, which is supposed to infer faster because of its smaller model size, consumes even more time and energy than ResNet, which indicates its inefficiency when deploying on RRAM crossbars.

\begin{figure}[!t]
\centering
\includegraphics[scale=0.4]{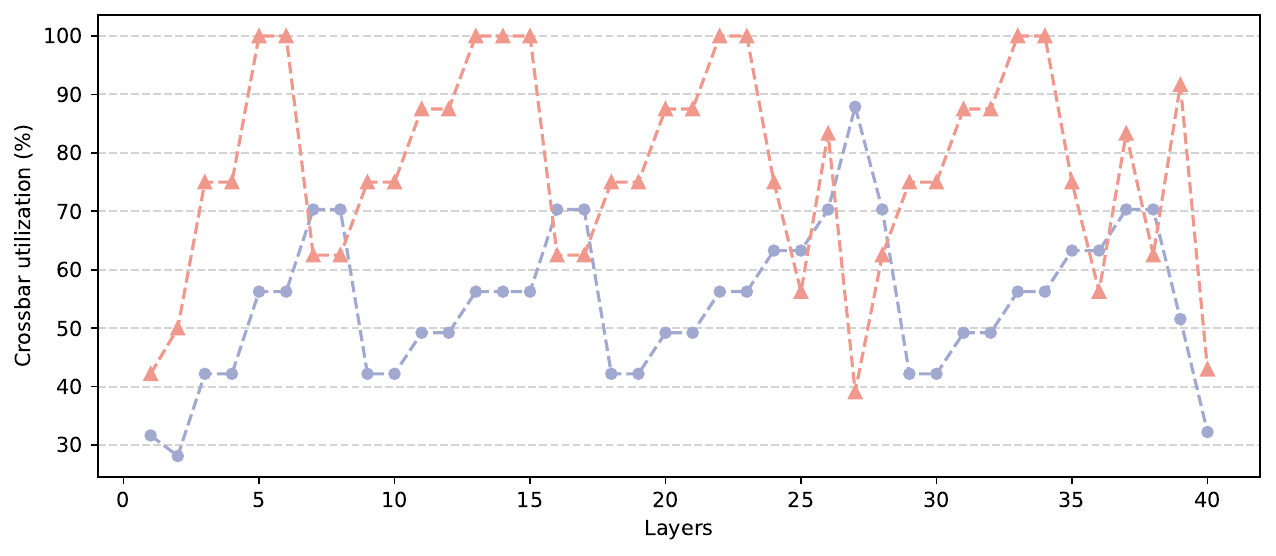}
\caption{Crossbar utilization, assuming $64 \times 64$ RRAM crossbars, of each layer for (red) DenseNet with GR=16 and (purple) DenseNet with GR=12.}
\label{fig:crossuti}
\vspace{-0.4cm}
\end{figure}

To this end, a detailed analysis of the inefficiency of DenseNet provides insights for efficient architectural (re)design: DenseNet consumes significant hardware resources due to a waste of on-chip resources. Specifically, Fig.~\ref{fig:crossuti} depicts layer-by-layer crossbar utilization of two DenseNets. It is clear that the utilization of many layers in these DenseNets is below 80\%. Here crossbar utilization refers to the ratio of RRAM cells programmed with network weights to the total number of RRAM cells. Cells that are not programmed with DNN weights still consume energy for meaningless computation, which significantly increases the hardware cost.

Fig.~\ref{fig:densenet} reveals the reason for the low crossbar utilization by taking the second layer in a dense module as an example. The linearly increasing input channels causes convolution kernels to extend in the channel dimension. Suppose the length of the first-layer inputs equals to the height of one crossbar, another crossbar is needed to deploy the second layer because of its extended channel dimension. This overshot length is shorter than that of the crossbar dimension, yielding unused cells marked by blank in the figure. 

Even more, the overall computation of DenseNet grows quadratically with respect to depth. This is why DenseNet introduces $1\times 1$ convolution as bottleneck layers before each $3\times 3$ convolution to reduce the growth of computation and improve computational efficiency. These additional $1\times 1$ convolutional layers increase the energy consumption and slow down inference. In short, the reason behind the inefficiency of both depthwise convolution and linearly increasing inputs is their failure to fully utilize the on-chip resource. Improving crossbar utilization is therefore critical for an efficient RRAM-friendly neural network.

\begin{figure}[!t]
\centering
\includegraphics[scale=0.5]{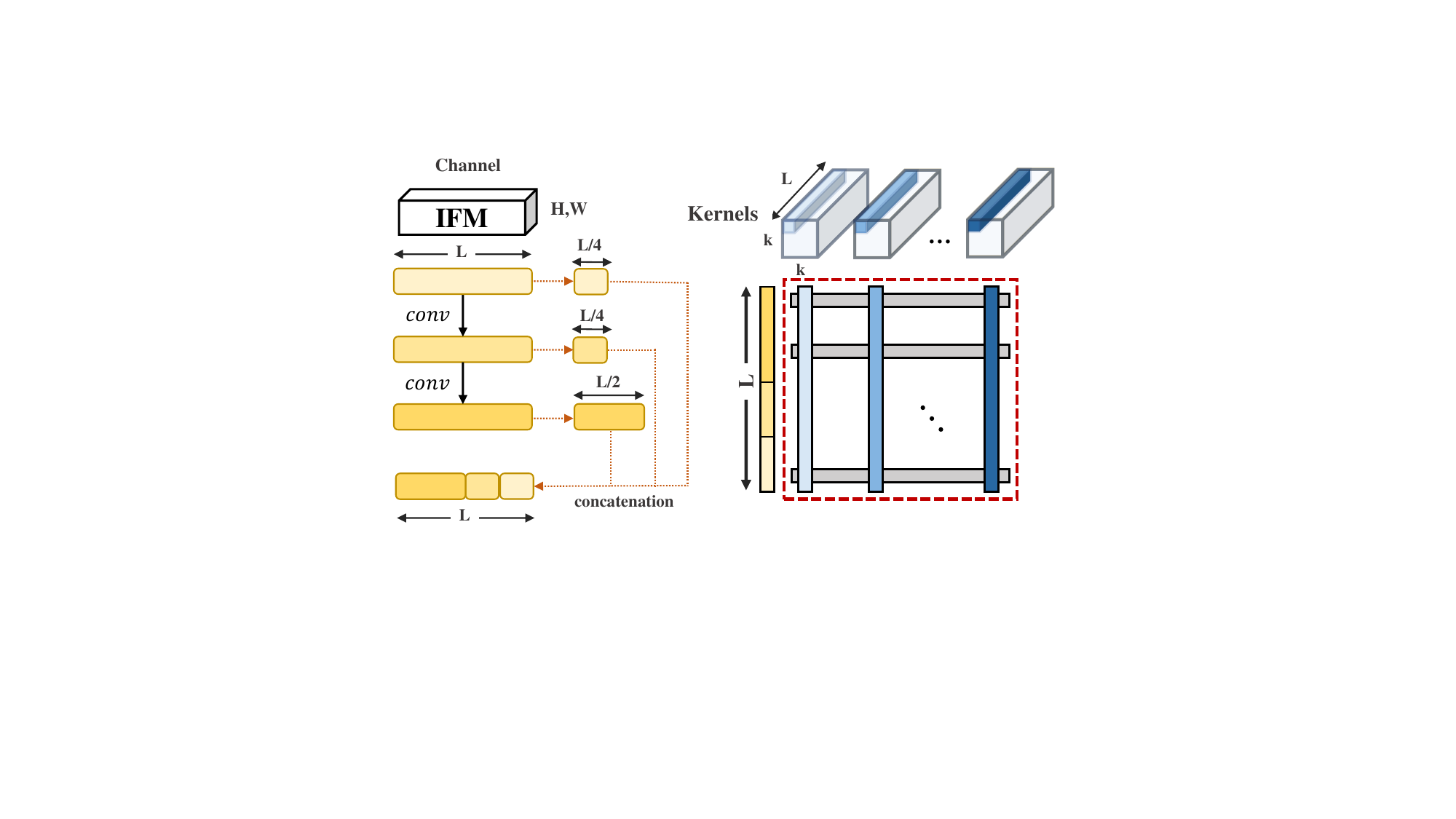}
\caption{(Left) Fixed-channel-length inputs in our module. (Right) Mapping the last layer in our module to RRAM crossbars.}
\label{fig:ours}
\vspace{-0.4cm}
\end{figure}

\section{Time- and Energy-Efficient CNN on RRAM}
\label{sec:methodology}
Section~\ref{sec:rethink} shows that dense connections that aggregate all intermediate layers induce inevitable inefficiency. Although such concatenation improves the quality of feature maps compared to summation,  the significant time and energy consumption when deploying DenseNet on RRAM crossbars render it sub-optimal. In this regard, we propose an RRAM-friendly efficient architecture that inherits the merit of dense connections while ensuring crossbar utilization. Fig.~\ref{fig:ours} depicts the proposed module from the perspective of channel dimension. In contrast to the dense block in DenseNet, the input size of each layer in our module remains the same. Instead of directly passing the output of the 3rd layer to the last layer as input, our module concatenates feature maps from the previous 3 layers to build the input for the last layer. Since the number of channels stays fixed across layers, there is no need to adopt additional $1\times 1$ convolutions for shrinking the channel dimension. This makes our module contain fewer layers than the original dense block. Moreover, we only concatenate feature maps once in each module, thereby accelerating the inference speed on RRAM chips compared to the original DenseNet. Fig.~\ref{fig:ours} shows the weight-to-RRAM cell mapping of the last layer in our module. The input size for all layers in our module remains the same, making it easy to prescribe the crossbar dimensions for practically full utilization. Following the convention~\cite{ResNet}, our model has four stages, each containing one module shown in Fig.~\ref{fig:ours}.

\begin{figure}[!t]
\centering
\includegraphics[scale=0.25]{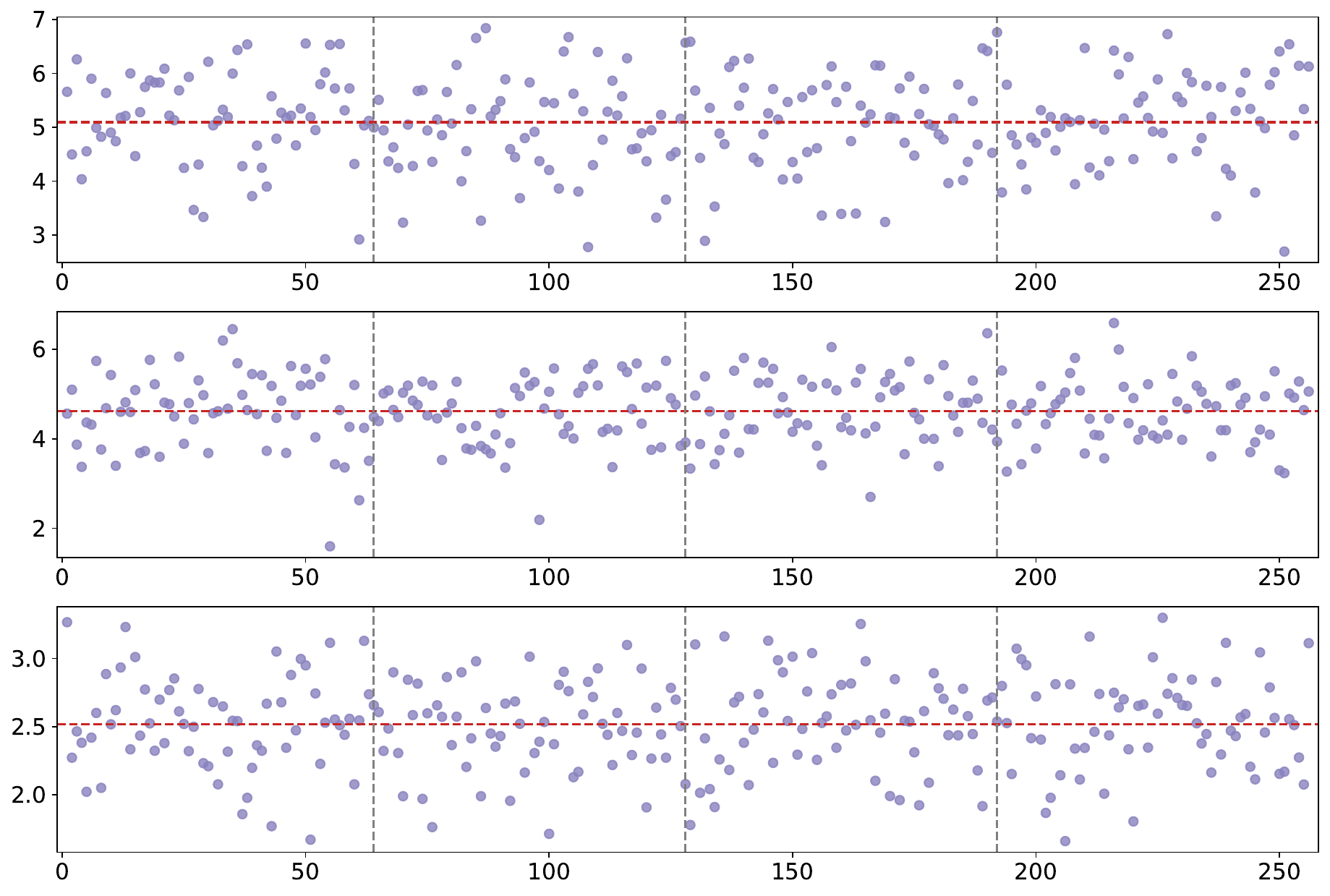}
\caption{Feature map ranks from different convolutional layers in our module on CIFAR-10. From top to bottom are the  rank plots of the first 3 layers corresponding to Fig.~\ref{fig:ours}. The x-axis represents the indices of feature maps and the y-axis denotes the averaged ranks. The mean rank in each quarter is further marked by dotted red lines.}
\label{fig:ranks}
\end{figure}

For simplistic circuit design and thus higher speed and lower power, we select the first quarter of feature maps in the 1st and 2nd layers and the first half of feature maps in the 3rd layer. HRank~\cite{hrank} proposes evaluating the importance of feature maps by their ranks. Fig.~\ref{fig:ranks} shows the ranks of individual quarters from different layers. The average ranks of different quarters in a layer turn out to be almost identical, showing their nearly equal importance. Table~\ref{tab:ablation_q} presents the ablation study on selecting different quarters/halves in our design, showing negligible difference with respect to the choices, which in turn justifies our simplistic circuit configuration.

\section{Experiments}
\label{sec:experiments}
\begin{figure}[!t]
\centering
\includegraphics[scale=0.34]{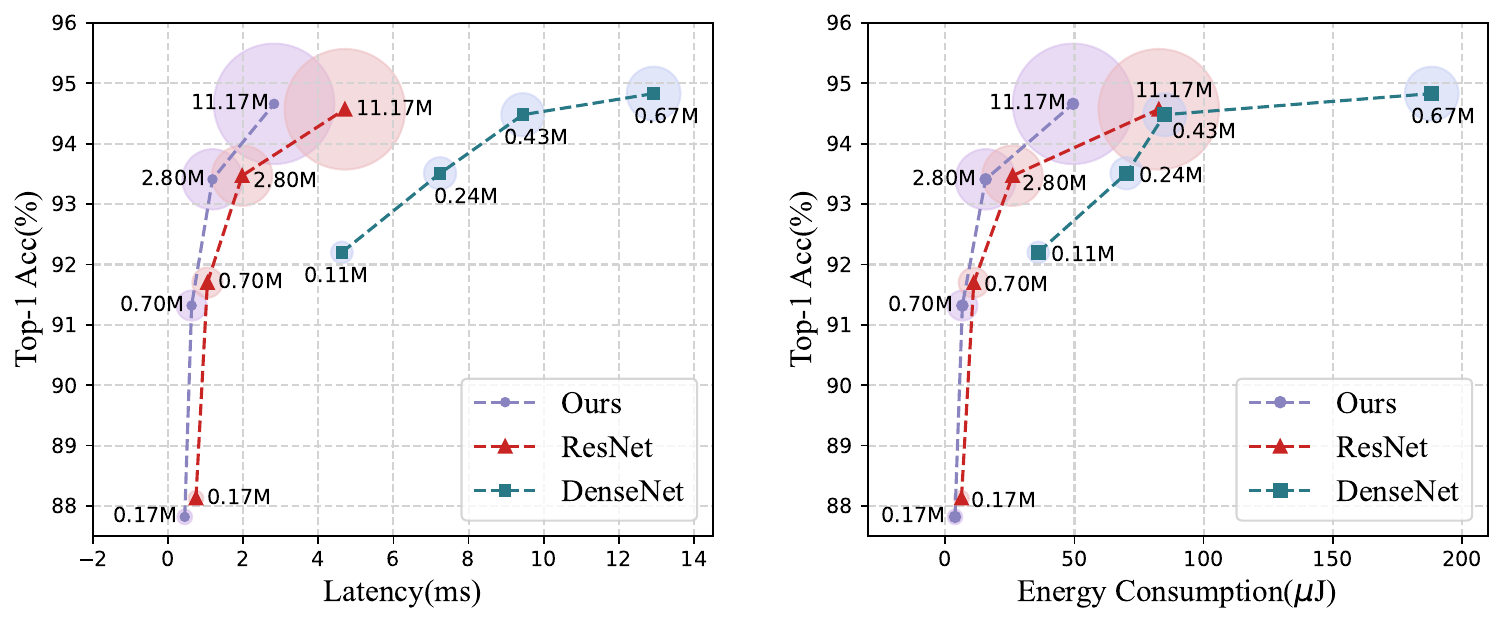}
\caption{Accuracy vs (left) latency and (right) energy consumption on different DNN architectures and sizes trained on the CIFAR-10 dataset and deployed onto RRAM crossbars.}
\label{fig:mainResult}
\end{figure}

The proposed architecture is trained and validated by PyTorch 1.7.0. All experiments are conducted on NVIDIA RTX3090 platform with a 32GB frame buffer. As for the RRAM simulation, the DNN+NeuroSim workflow is adopted to evaluate various metrics at the circuit level. The crossbar size is set to $64 \times 64$, where the results generalize to other common sizes. Fig.~\ref{fig:mainResult} shows the latency, energy consumption and accuracy on the CIFAR-10 dataset of different models. To better visualize the trends, we scale the width (viz. number of channels) to obtain different model sizes. DenseNet outperforms other models when the width is small in terms of accuracy, but all models show comparable performance when the width increases. Our model consumes less time and energy to achieve the same accuracy. Compared to ResNet, our model does not employ residual connections, thus reducing hardware costs. Indeed, we found the residual connections in ResNet contribute $\sim$10\% of the latency and energy consumption.

\begin{figure}[!t] 
\centering
\includegraphics[scale=0.3]{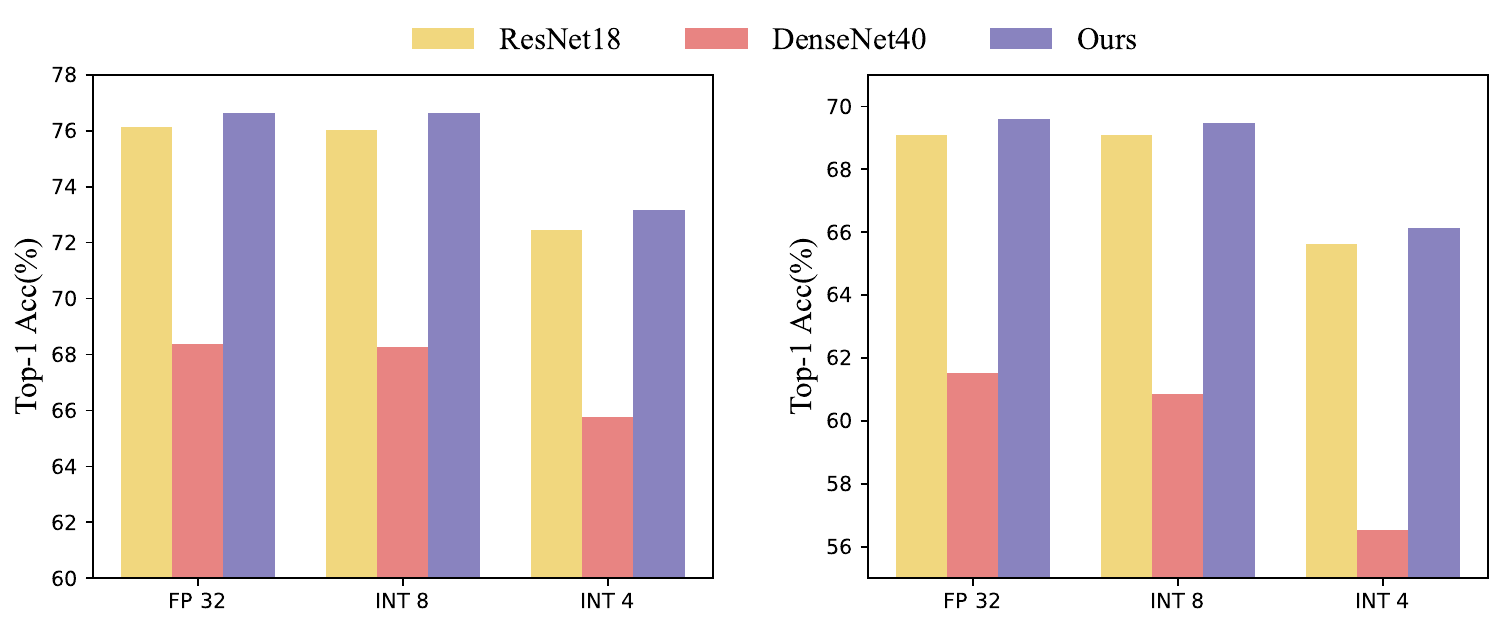}
\caption{The comparison between ResNet, DenseNet and our model in terms of accuracy on (left) CIFAR-100 and (right) ImageNet datasets after INT8 and INT4 quantization.}
\label{fig:quant}
\end{figure}

Table~\ref{tab:acc} lists the performance of ResNet, DenseNet and our model on the ImageNet dataset. Fig.~\ref{fig:quant} shows the performance of different quantized models on both CIFAR-100 and ImageNet datasets. Both INT8 and INT4 quantizations adopt the approach in~\cite{DyBit}. To maintain consistency, we selected the DenseNet with the largest width (0.67M) in Fig.~\ref{fig:mainResult} for accuracy analysis on the ImageNet dataset.

\begin{table}[!t]
\centering{
\setlength{\tabcolsep}{4mm}{
\begin{tabular}{cccc}
\toprule
\multicolumn{3}{c}{\textbf{Portions pass to the last layer}} & \multirow{2}{*}{\textbf{Top-1 Acc (\%)}} \\ \cline{1-3}
\textbf{Layer 1}  & \textbf{Layer 2} & \textbf{Layer 3} &                                     \\ \hline
1/4 (F)                 & 1/4 (F)                  & 1/2 (F)                  & \textbf{91.32}                             \\
1/4 (L)                 & 1/4 (L)                & 1/2 (L)                 & 91.29                          \\      
1/4 (M)                 & 1/4 (M)                & 1/2 (M)                & 91.31                             \\

\bottomrule                        
\end{tabular}}
}
\caption{Ablation study on concatenating different parts of feature maps in the first 3 layers to the last layer on CIFAR-10 dataset. F: first (quarter/half), M: middle part, L: (last quarter/second half).}
\label{tab:ablation_q}
\vspace{-0.4cm}
\end{table}

Table~\ref{tab:ablation} shows the ablation study of our module on CIFAR-10. We change the combinations of the input of the last layer. Following the common practice~\cite{ResNet}, we set the number of channels in four stages to be [16, 32, 64, 128]. From the table, decreasing the portion of feature maps from the 3rd layer brings increments in the beginning and then leads to drops in terms of accuracy, whereas our design achieves peak accuracy among all modules.

\begin{table}[!t]
\centering{
\setlength{\tabcolsep}{4mm}{
\begin{tabular}{ccc}
\toprule
\textbf{Model}       & \textbf{Top-1 Acc(\%)} & \textbf{Top-5 Acc(\%)}      \\ \hline
VGG-11 \cite{VGG}         & 69.02              & 88.62                  \\
ResNet-18 \cite{ResNet}         & 69.75              & 89.01                  \\
DenseNet-40 \cite{DenseNet}             & 61.52              & 82.47              \\
\textbf{Ours}                              & \textbf{69.83}              & \textbf{89.52}    \\ \bottomrule
\end{tabular}}
}
\caption{The comparison between traditional CNN models and our model on the ImageNet dataset.}
\label{tab:acc}
\end{table}

\begin{table}[!t]
\centering{
\setlength{\tabcolsep}{4mm}{
\begin{tabular}{cccc}
\toprule
\multicolumn{3}{c}{\textbf{Portions passed to the last layer}} & \multirow{2}{*}{\textbf{Top-1 Acc (\%)}} \\ \cline{1-3}
\textbf{Layer 1}  & \textbf{Layer 2} & \textbf{Layer 3} &                                     \\ \hline
1/16 (F)                 &1/16 (F)                 & 7/8 (F)                & 90.86                             \\
1/8 (F)                 &1/8 (F)                 &3/4 (F)                 & 90.88                          \\
\textbf{1/4 (F)}                 &\textbf{ 1/4(F) }               & \textbf{1/2 (F)}                & \textbf{91.32}                             \\
1/4 (F)                 & 1/2 (F)                & 1/4 (F)                & 91.03                             \\
1/2 (F)                 & 1/4 (F)                & 1/4 (F)                & 90.81     \\
1/8 (F)                 & 3/4 (F)                & 1/8 (F)                & 90.92                             \\
3/4 (F)                 & 1/8 (F)               & 1/8 (F)                & 90.75     \\
1/16 (F)                 & 7/8 (F)                & 1/16 (F)                & 90.71                             \\
7/8 (F)                 & 1/16 (F)               & 1/16 (F)                & 90.60     \\
\bottomrule                        
\end{tabular}}
}
\caption{Ablation study on concatenating different percentages of feature maps in the first 3 layers to the last layer in CIFAR-10 training. F: first 1/16 (or 1/8, 1/4, 1/2) channels.}
\label{tab:ablation}
\vspace{-0.4cm}
\end{table}

\section{Conclusion}
\label{sec:conclusion}  
RRAM chips show promising prospects in accelerating DNN inference. However, existing lightweight CNNs, such as MobileNet, lead to low crossbar utilization and thus high power and latency. This paper first identifies the inefficiency of depthwise convolution arising from the mapping method. Simulation on the hardware cost of DenseNet then reveals the large hardware consumption is caused by its low crossbar utilization. With these insights, we propose an RRAM-friendly network that inherits the merit of dense connection and low parametric count. Our module concatenates feature maps of the initial layers to form the input of the last layer, keeping the input sizes of each layer to be the same. Experiments show that our model exhibits lower latency and energy than conventional ResNet and DenseNet, while up-keeping performance.

\section{Acknowledgement}
\label{sec:acknowlege}
This work was supported by the Mainland-Hong Kong Joint Funding Scheme Project MHP/066/20, the Theme-based Research Scheme (TRS) project T45-701/22-R, and in part by ACCESS – AI Chip Center for Emerging Smart Systems, sponsored by InnoHK funding, Hong Kong SAR.

\tiny
\bibliographystyle{ieeetr}
\bibliography{bare_conf}

\end{document}